\documentclass
[twocolumn,aps,prc,amsmath,showpacs,amssymb,floatfix]
{revtex4}
\usepackage{amssymb}

\usepackage{CJK}                     
\usepackage[dvips]{graphicx}
\usepackage{bm}                      
\usepackage{mathptmx}                
\usepackage{dcolumn}                 
\usepackage{array}
\def\eps{\epsilon}
\begin{document}

\title{Massive hybrid stars with a first order phase transition}
\author{A. Li$^{1,2}$\footnote{liang@xmu.edu.cn}, W. Zuo$^{2,3}$, G.-X. Peng$^{4,5}$}
\affiliation{$^1$ Department of Astronomy and Institute of
Theoretical Physics
and Astrophysics, Xiamen University, Xiamen, Fujian 361005, China\\
$^2$ State Key Laboratory of Theoretical Physics, Institute of
Theoretical Physics, Chinese Academy of Sciences, Beijing 100190,
China\\
$^3$ Institute of Modern Physics, Chinese Academy of Sciences,
Lanzhou 730000, China\\
$^4$ School of Physics, University of Chinese Academy of Sciences, Beijing 100049, China\\
$^5$ Theoretical Physics Center for Science Facilities, Institute of High Energy Physics, Beijing 100049, China\\
}
\date{\today}

\begin{abstract}
  We develop our previous study of the transition to deconfined quark phase in neutron stars,
  including the interaction in the quark equation of state to the leading order in the  perturbative expansion within
   the confinement density-dependent mass model.  Using the Gibbs conditions
   the hadron-quark mixed phase is constructed matching the latter with the hadron equation of state
   derived from the microscopic Brueckner-Hartree-Fock approximation. The influence of quark
interaction parameters on threshold
    properties and phase diagram of dense neutron star matter are discussed in detail. We find that
    the leading-order quark interaction expands the
     density range of the mixed phase, pushing forward the disappearance of the hadron
     phase. Moreover, since the equation of state could turn out to be stiffer, a
     high-mass hybrid star is possible with mixed-phase core with typical parameter sets.
\end{abstract}


\maketitle
\section{Introduction}

With the complementary investigations on  heavy-ion collective
flows and two precise measurements of heavy mass
pulsars~\cite{2times}, the study of neutron stars (NSs) has become
a more and more active field of research. \emph{Ab initio} lattice
QCD simulations and planned missions LOFT~\cite{LOFT} and NICER
will certainly promote further our current understanding of the
underlying baryonic forces, high-density equation of state (EoS),
and NSs' core properties~\cite{rep}. There might be three kinds of
non-nucleonic components in the NS interior:  free
quarks~\cite{Li08q}, mesons~\cite{Li10k}, and
hyperons~\cite{Li11y,Li14y}). But no \emph{ab initio} calculations
are available so far for their relevance and abundance in NS,
because such calculations is unachievable due to the complicated
nonlinear and nonperturbative nature of QCD.

In the previous article~\cite{Li08q} we have investigated the NS
structure within the Brueckner-Hartree-Fock (BHF) approximation,
which is currently one of the most advanced microscopic approaches
to the EoS of nuclear matter~\cite{book}. In that paper BHF was
combined with the confinement density-dependent mass model
(CDDM)~\cite{cubic} for the quark phase to model hybrid stars
(HSs), limiting ourselves to include only the confinement
potential in the quark mass scaling. In this work we further
extend our calculations by including also short-range
leading-order perturbative interactions in the employed quark
matter EoS model (i.e., a new version of CDDM
model~\cite{horder}), and explore the consequences for HS
structure.

Although in the high temperature region hadron-quark phase
transition is crossover as explored by lattice QCD
simulations~\cite{Aoki06}, the order of phase transition at zero
temperature is still an open problem, as well as the existence of
a critical end point (CEP) in the QCD phase diagram. We assume in
the present work that the hadron-quark phase transition in cold
NSs is a first-order one, and use the Gibbs construction to match
the hadron EoS and the quark EoS for obtaining the mixed phase (see Refs.~\cite{12,13} for more discussions on its dependencies on physical situations).
That is, the pressure is taken to be the same in the hadron-quark
mixed phase to ensure mechanical stability, and would increase
monotonically with baryon chemical potential. In the mean time, a
global charge neutrality is assumed. Other authors used a smooth
crossover~\cite{Bla14,Hat13} to obtain the transition.

We provide a short overview of the theoretical framework and
discussions of our results in Sect.~II, before drawing conclusions
in Sect.~III.

\section{Formalism and Discussion}

\subsection{The hadron phase}
Let us first address the hadron phase, that is nuclear matter
consisting of nucleons in $\beta$-equilibrium with electrons:
\begin{eqnarray}
n    \rightleftharpoons   p+e^- + \bar{\nu}_e \cdot
\end{eqnarray}
We omit muons in the following calculations since they are
irrelevant to the purpose of the present work. Under the condition
of neutrino escape, this equilibrium can be expressed as
\begin{eqnarray}
\mu_n - \mu_p = \mu_{e},  \label{mu_N}
\end{eqnarray}
And the requirement of charge neutrality implies
\begin{eqnarray}
n_p = n_{e^-}, \label{neutral}
\end{eqnarray}
where $\mu_i$ ($n_i$) is the chemical potential (the number
density) of component $i$.

The chemical potentials of the non-interacting electrons are
obtained by solving numerically the free Fermi gas model. The
nucleonic chemical potentials required in Eq.~(\ref{mu_N}) are
derived from the energy density of nuclear matter, based on the
BHF nuclear many-body approach described
elsewhere~\cite{bhf}. Here the input bare nucleon force we
employed is the Argonne V18 two-body interaction~\cite{v18},
accompanied by a microscopic three body force constructed from the
meson-exchange current approach~\cite{Zuowei02prc}. The
corresponding nuclear EoS reproduces correctly the nuclear matter
saturation point and fulfills several requirements from the
nuclear phenomenology~\cite{bbb}. We mention here that the model developed in this article misses some important aspects, such as the inclusion of hyperons. The interplay between hyperons and free quarks is quite important and deserves additional investigation, especially when more reliable empirical inputs will be available, especially on the hyperon-nucleon and hyperon-hyperon interaction.

Once the nuclear EoS and the nucleonic chemical potentials of
nuclear matter are known, one can then proceed to calculate the
composition of the hot $\beta$-equilibrium matter by solving Eqs.
(\ref{mu_N}) and (\ref{neutral}), together with the conservation
of the baryon number, $n_n+n_p = n_{\mathrm{B}}$. Finally the
total energy density $\eps_N$ and the total pressure $p_N$ of the
system are obtained after adding the standard contribution of
electrons.

\subsection{The quark phase}

The quark phase is considered as a mixture of interacting $u$,
$d$, $s$ quarks  in $\beta$-equilibrium with electrons:
\begin{eqnarray}
d    &\rightleftharpoons&   u+e^- + \bar{\nu}_e ,\\
s   &\rightleftharpoons&   u+e^- + \bar{\nu}_e ,\\
s + u    &\rightleftharpoons&   d + u .
\end{eqnarray}

The crucial problem in studying the quark matter is to treat the quark confinement in a proper way. In the framework of the bag model, an extra constant, the famous bag constant $B$, is introduced which provides a negative pressure to confine quarks within a finite volume. That is, the quark mass is infinitely large outside the bag, and finite and constant within the bag. As is well known, however, particle masses vary from the vacuum to a medium. Taking advantage of the density dependence, one can describe quark confinement without using the bag constant. Instead, the quark confinement is achieved by the density dependence of the quark masses derived from in-medium chiral condensates~\cite{cubic}. That is the CDDM model we employed in the present study. A large amount of investigation have been performed in the framework of this model, and it has been developed greatly in recent years (see Ref.~\cite{horder} and references therein).

In the employed CDDM model, strong interactions between quarks are
mimicked by an equivalent mass to be determined:
\begin{eqnarray}
H_{\mathrm{QCD}}&=&H_{\mathrm{k}}+\sum_{q=u,d,s}m_{q0}\bar{q}q
               +H_{\mathrm{I}}\\ \nonumber
               &\equiv&H_{\mathrm{k}}+\sum\limits_{q=u,d,s}m_q\bar{q}q
\end{eqnarray}
where $m_{q0}$ ($q = u, d, s$) are the quark current mass,
$H_{\mathrm{k}}$ is the kinetic term, $H_{\mathrm{I}}$ is the
interacting part. The equivalent mass $m_q$ embodies all the
interaction effects between quarks. That is, the contributions
from both the scalar field and the Lorentz vector field can be
included in this way~\cite{both}.

There are several ways to determine the equivalent mass in the
literature (see \cite{liangmn} and references therein). Here we
use a recently derived mass formula at zero
temperature~\cite{horder}:
\begin{eqnarray}
m_q \equiv m_{q0}+ m_{\mathrm{I}}=m_{q0} +
\frac{D}{n_{\mathrm{B}}^{1/3}} + Cn_{\mathrm{B}}^{1/3}.
\label{mqT0}
\end{eqnarray}
where $m_{\mathrm{I}}$ is the interacting mass, parameterized as a
function of the baryon number density $n_{\mathrm{B}}$. $D$ term is derived from
the non-perturbative linear confinement of quarks (see also
\cite{cubic}), and $C$ term comes from the short-range leading
contribution of perturbative interactions. These two terms correspond to the two leading terms in both directions when
expanding the equivalent mass to a Laurant series of the holistic
Fermi momentum, respectively~\cite{horder}. The confinement
interaction dominates at lower densities, while the perturbative
interactions becomes more important at higher densities. This model gives
reasonable results for the sound velocity. Also, strange quark
matter in bulk still has the possibility of absolute stability for
a wide range of parameters. As to the quark current mass, in our
calculations we take $m_{u0}=m_{d0}=0$ and $m_{s0}=95$ MeV.

The parameter $D$ has a lower bound $D^{1/2} = 156$ MeV, and an
upper bound $D^{1/2} = 270$ MeV~\cite{WenxjPRC72}. The lower bound
comes from the nuclear physics constraint, demanding that at
$P=0$, non-strange nuclear matter should be stable against decay
to $(ud)$ quark matter. This leads to the condition $E/A >
M_{^{56}\mbox{Fe}}c^2/56 = 930$ MeV for $(ud)$ quark matter, which gives
the above mentioned lower bound. The upper bound can be derived
from a relation between $D$ and the quark-condensate and the known
range of values for this condensate~\cite{WenxjPRC72}. The upper boundary of $270$
MeV is in fact a very conservative one. According to the updated
quark condensate determined nowadays very precisely by lattice
QCD~\cite{upqcd}, a range of ($161$ MeV, $195$ MeV) can be
obtained. Therefore in this work, we take two typical values of
the confinement parameter as $D^{1/2} = 170$ MeV, $190$ MeV as
inferred by the newest lattice QCD results~\cite{upqcd}.

The parameter $C$ depends on how the strong coupling runs and it
is determined so to have a upper bound of $C =
1.1676$~\cite{horder}. Previous calculations~\cite{horder} of pure
quark stars employing a CDDM EoS lead to a maximum mass as high
as $2M_{\odot}$ with a parameter set ($C, D^{1/2}) = (0.7, 129$
MeV). We will then employ $C = 0.7$ to perform calculations and
change its value in a certain range for comparison.

The relevant chemical potentials $\mu_u$, $\mu_d$, $\mu_s$, and
$\mu_e$ satisfy the weak-equilibrium condition (we again assume
neutrino escape):
\begin{eqnarray}
 \mu_u- \mu_d = \mu_{e}, ~~~\mu_d=\mu_s{\bf .} \label{weak}
\end{eqnarray}
The baryon number density and the charge density can be written as
\begin{eqnarray} \label{qmeq3}
n_{\mathrm{B}}=\frac{1}{3}(n_u+n_d+n_s){\bf ,}
\end{eqnarray}
\begin{eqnarray} \label{qmeq4}
q_Q=\frac{2}{3}n_u-\frac{1}{3}n_d-\frac{1}{3}n_s-n_{e}.
\end{eqnarray}
The charge neutrality condition requires $q_Q=0$.

\begin{figure}
\centering
\includegraphics[width=8.5cm]{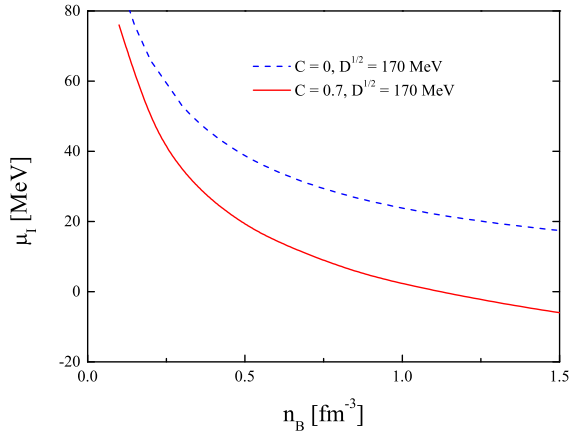}
\caption{(Color online) Extra chemical potential as a function
of the baryon density, for both $C = 0$ and $C = 0.7$ cases with
$D^{1/2} = 170$ MeV.}\label{fig1}
\end{figure}

Since the quark masses $m_i$ are density dependent, the quark
chemical potentials $\mu_i$ have an additional term
$\mu_{\mathrm{I}}$ with respect to the free Fermi gas model ($j =
u, d, s$):
\begin{eqnarray}
&&\mu_i = \frac{\partial \eps_i}{\partial \nu_i}
\frac{\mathrm{d}\nu_i}{\mathrm{d} n_i}
 +\sum_j \frac{\partial \eps}{\partial m_j}\frac{\partial m_j}{\partial n_i}
= \sqrt{\nu_i^2+m_i^2}-\mu_{\mathrm{I}} \label{quasi}
\end{eqnarray}
The quark energy densities are
\begin{eqnarray}
&&\eps_i =\frac{3}{\pi^2}\int_0^{\nu_i}\sqrt{p^2+m_i^2}\,p^2\,\mbox{d}p, \\
&&\eps =\sum_i\eps_i,
\end{eqnarray}
where $\nu_i=(\pi^2 n_i)^{1/3}$ are the Fermi momenta and
$\partial m_j/{\partial n_i}$ are derived from Eq.~(\ref{mqT0}) by
taking the derivative of the baryon density. The quark pressure is
calculated as $p = -\eps + \sum_i \mu_i n_i$. Solving Eqs.
(\ref{weak}), (\ref{qmeq3}) and (\ref{qmeq4}), the total energy
density $\eps_q$ and pressure $p_q$ of the system can be obtained
after adding the contribution of the leptons.

\begin{figure}
\centering
\includegraphics[width=8.5cm]{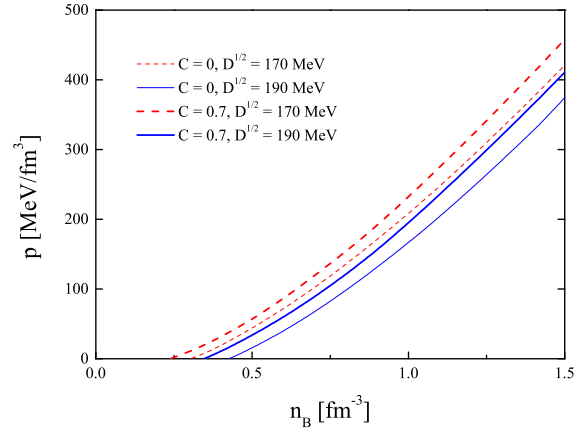}
\caption{(Color online) Pressure of $\beta$-stable quark matter as
a function of baryon density for two values of $C = 0, 0.7$, and
two values of $D^{1/2} = 170$ MeV, $190$ MeV.}\label{fig2}
\end{figure}
In our previous calculations~\cite{Li08q} there is only the term
$D$ in the quark mass scaling, which results in relatively low
quark thresholds and small HS masses, similar to the
calculations~\cite{hybridbhf04} within the color dielectric model
and the MIT bag model. As will be shown later, inclusion of the
$C$ term may bring strongly repulsive quark interactions, pushing
the quark matter to appear at appropriately high densities, and
consequently making massive HSs in the model.

For a better understanding of this point, we present in
Fig.~\ref{fig1} the modification of the extra chemical potentials
$\mu_{\mathrm{I}}$ induced by the nonzero perturbative parameter
$C$. The calculations are done with fixed $D^{1/2} = 170$ MeV and
two values of $C = 0, 0.7$. We see that with increasing density, $\mu_{\mathrm{I}}$ always decreases. A nonzero $C$ makes it
decrease faster with the density. The decrease is even more
pronounced at higher densities, because the term
$Cn_{\mathrm{B}}^{1/3}$ is an increasing function of the density.
Also, the extra chemical potential $\mu_{\mathrm{I}}$ changes from
positive to negative values at high densities for $C = 0.7$,
indicating that the  term $C$ brings repulsions, and quarks in this case are more strongly
interacting with each other. This effect of $C$ is opposite to
that of $D$, since the latter arises from the quark confinement
potential, and the increase of this parameter will bring
attractions and soften the EoSs of the matter. Fig.~\ref{fig2}
clearly demonstrates these effects, where the EoSs of
$\beta$-stable quark matter are shown for the two values of $C =
0,~0.7$, and the two values of $D^{1/2} = 170$ MeV, $190$ MeV. The
increase of the confinement parameter $D$ will soften the EoS,
while the perturbative parameter $C$ will stiffen it. The repulsive
nature of the  term $C$ will have crucial consequences for the
structure of the resulting HSs as seen later.

\subsection{The mixed phase}
\begin{figure}
\centering
\includegraphics[width=8.5cm]{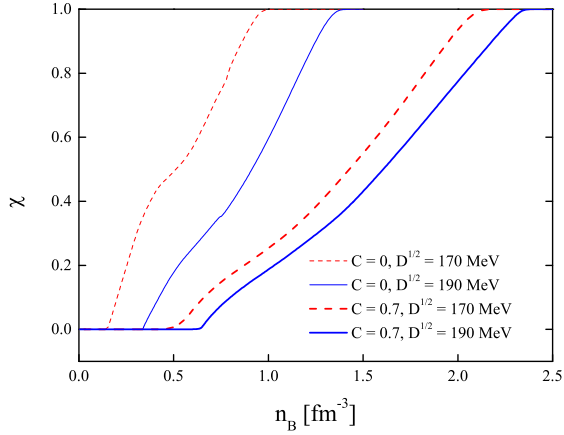}
\caption{(Color online) Quark fraction $\chi$ as a function of the
baryon density, for two values of $C = 0, 0.7$, and two values of
$D^{1/2} = 170$ MeV, $190$ MeV. }\label{fig3}
\end{figure}

Let us consider the mixed phase made of nucleon matter in
equilibrium with a gas of $u, d, s$ quarks and electrons. Assuming
the global charge conservation, the conservation laws can be
imposed introducing the quark fraction $\chi$ defined as
\begin{equation}
\chi\equiv V_{\mathrm{q}}/V,
\end{equation}
where $V$ is the total volume and $V_{\mathrm{q}}$ the volume
occupied by quarks. In terms of $\chi$ the total baryon density,
total electric charge and total energy density are written
\begin{equation} \label{rhotot}
n_{\mathrm{B}}
 = (1-\chi) n_{\mathrm{N}} +\chi n_{\mathrm{q}},
\end{equation}
\begin{equation} \label{Qtot}
Q_{\mathrm{t}}
 = (1-\chi) Q_{\mathrm{N}} +\chi Q_{\mathrm{q}},
\end{equation}
\begin{equation} \label{Etot}
E_{\mathrm{t}} = (1-\chi) E_{\mathrm{N}} +\chi E_{\mathrm{q}},
\end{equation}
respectively. The quantities $n_{\mathrm{N}}$($n_{\mathrm{q}}$),
$Q_{\mathrm{N}}$( $Q_{\mathrm{q}}$), and $E_{\mathrm{N}}$
($E_{\mathrm{q}}$) are nucleonic (quark) number density, charge
density, and energy density, respectively.

Nucleonic chemical potentials are connected to quark chemical
potentials as follows:
\begin{eqnarray}
\mu_{\mathrm{n}}&=&\mu_u+2\mu_d,\\
\mu_{\mathrm{p}}&=&2\mu_u+\mu_d.
\end{eqnarray}
Therefore, there are only two independent chemical potentials. For
a given total density $n_{\mathrm{B}}$, the two independent
chemical potentials and the quark fraction $\chi$ can be
determined by solving the charge neutrality equation
$Q_{\mathrm{t}} = 0$ and the pressure balance equation
$p_{\mathrm{N}} = p_{\mathrm{q}}$.

\begin{figure}
\centering
\includegraphics[width=0.5\textwidth]{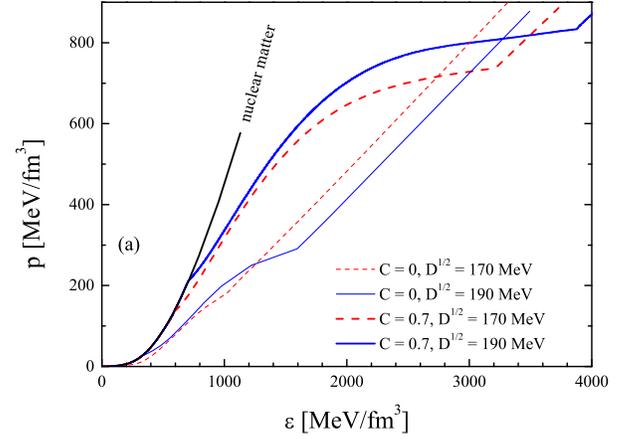}
\includegraphics[width=0.5\textwidth]{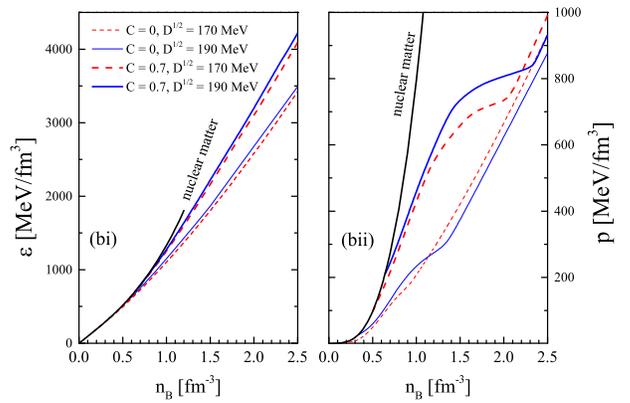}
\caption{(Color online) (Upper panel) Pressure of HS matter as a function of energy density and (lower panel) energy density and pressure of HS matter as a function of baryon density, for two values of $C = 0, 0.7$, and two values of $D^{1/2} = 170$ MeV, $190$ MeV. The results of pure nuclear matter are also shown for comparison.}\label{fig4}
\end{figure}

\begin{figure}
\centering
\includegraphics[width=0.5\textwidth]{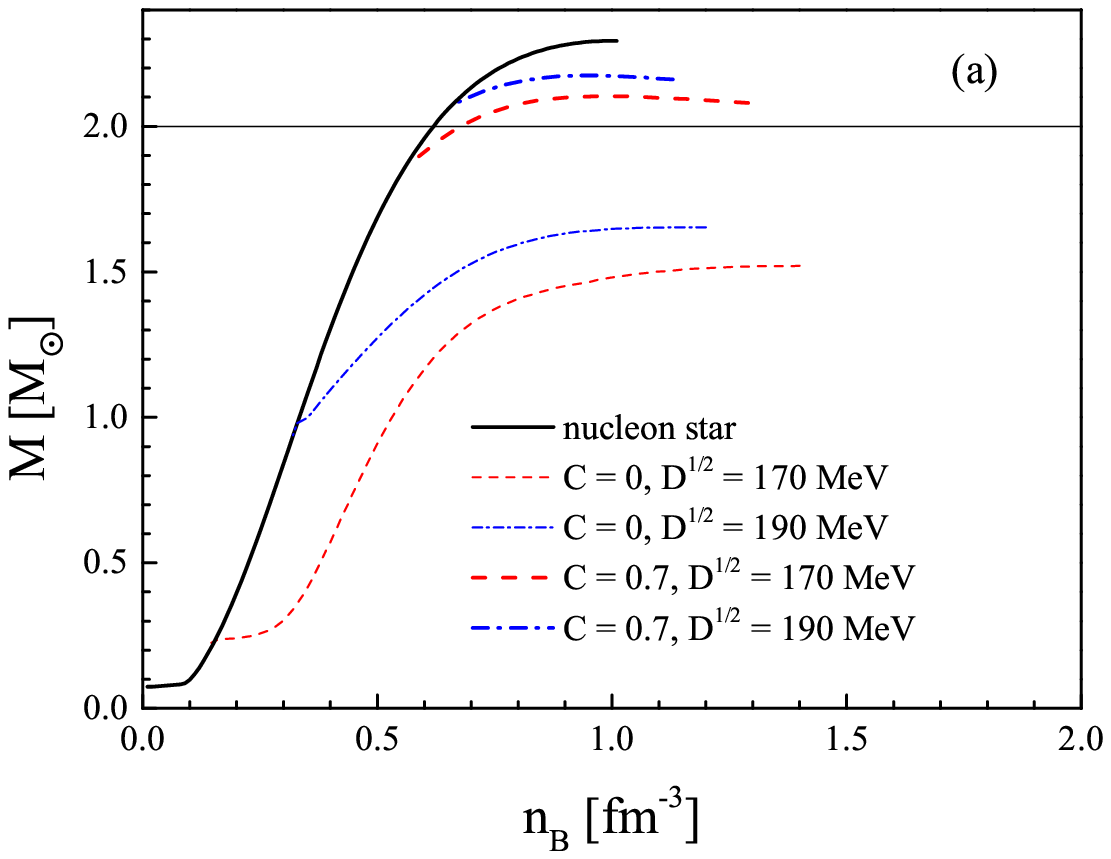}
\includegraphics[width=0.5\textwidth]{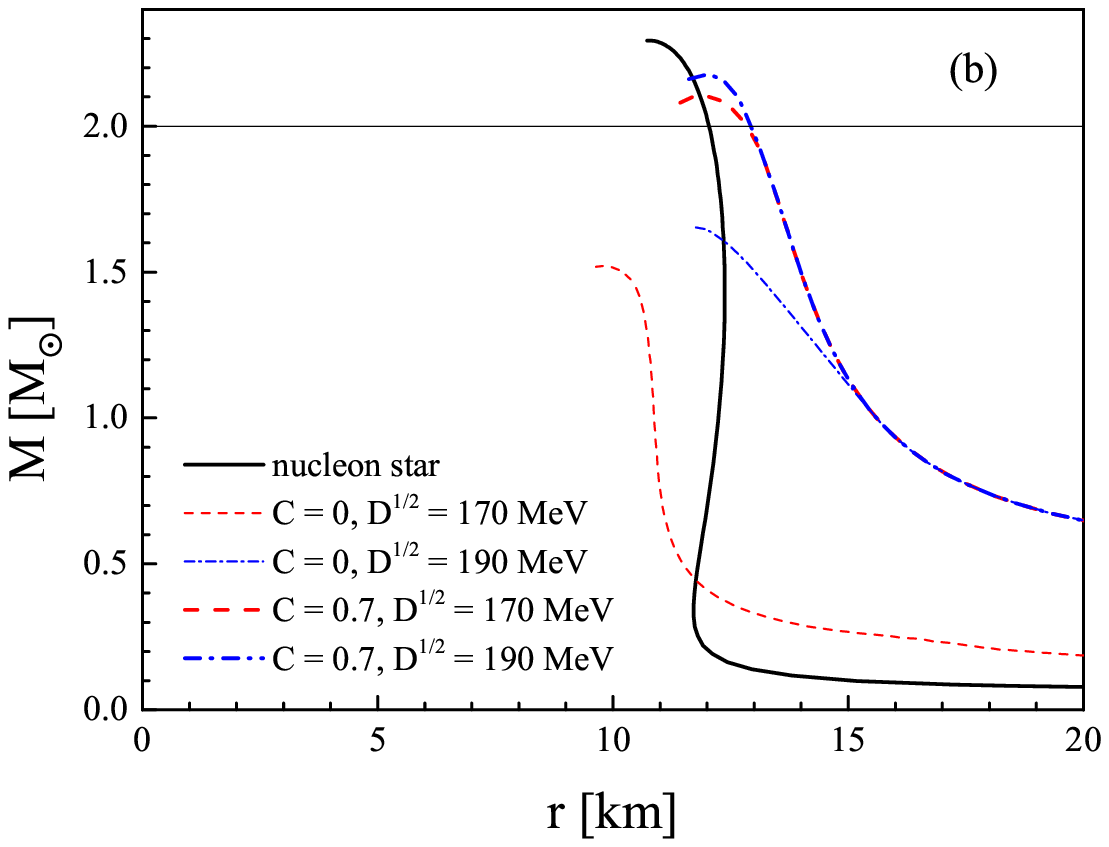}
\caption{(Color online) (Upper panel)
HSs' masses as a function of their central densities and (lower panel) HS mass-radius relations, for two values of $C = 0, 0.7$, and two values of $D^{1/2} = 170$ MeV, $190$ MeV.
The results of the nucleon star are also shown for comparison, as well as the recent 2-solar-mass constraint from the mass
measurements of PSR J1614-2230 and PSR J0348+0432~\cite{2times}.}\label{fig5}
\end{figure}
In Fig.~\ref{fig3}, the quark fraction $\chi$ is plotted as a
function of the baryon density, for the two values of $C = 0,
0.7$, and the two values of $D^{1/2} = 170$ MeV, $190$ MeV. As
already known from our previous work, the larger $D$ value pushes
the threshold of quark matter phase to higher densities in the
case of $C = 0$. This conclusion still holds when we adopt a
nonzero value of $C$  in the quark mass scaling, increasing the
critical density from $0.57$ fm$^{-3}$ to $0.64$ fm$^{-3}$ at ($C,
D^{1/2}) = (0.7, 190$ MeV). Also, a nonzero positive $C$ is found
to have the same effect, when the $C = 0.7$ cases are compared
with the corresponding $C = 0$ cases. A critical density $0.16$
fm$^{-3}$ ($0.33$ fm$^{-3}$) with $C = 0, D^{1/2} = 170$ MeV
($190$ MeV) is increased to $0.57$ fm$^{-3}$ ($0.64$ fm$^{-3}$)
with $C = 0.7, D^{1/2} = 170$ MeV ($190$ MeV).

In turn, when the $C$ value is chosen to be negative, that is
allowed by our model, quarks appear even earlier than the $C = 0$
case. This means that for typical model parameters free quarks
could be present below the nuclear saturation density $0.16$
fm$^{-3}$, which is unphysical, so that we have to exclude
negative $C$ values in the present work. Actually, a previous
study showed that negative $C$ values were allowed, for a simple
inclusion of one-gluon-exchange interaction between quarks and
degeneration 4 with respect to spin and flavor~\cite{1glue}. This
can be understood as follows: Since we start from the quark
potential, the spin and flavor degrees of freedom should be very
important. Especially, the spin-spin interaction between the
quarks plays an important role in calculating the effective
repulsion~\cite{spinspin}. A more realistic inclusion of the quark
potential as done in~\cite{string} could then lead to a more
reasonable result of the critical density around 0.5 fm$^{-3}$.
Furthermore, when the term  $C$ is included, together with an
upper quark threshold, the density range of the mixed hadron-quark
phase is expanded (around twice with the chosen parameters). This should be a general result when the Gibbs construction
is employed to achieve a first-order phase transition.

The stable configurations of a NS can be obtained from the
well-known hydrostatic equilibrium equations of Tolman,
Oppenheimer, and Volkov~\cite{shapiro}. At variance with pure
nuclear or quark stars, a HS may contain pure quark matter in the
core, pure nuclear matter near the outer part, and, in between, a
mixed phase of the quark and nuclear matter. We then employ
corresponding EoS models described above. They are shown in Fig.~\ref{fig4}, where the pressures of HS matter as a function of energy density are shown in the upper panel, and the energy densities and pressures of HS matter as a function of baryon density are shown in the lower panel, for two values of $C = 0, 0.7$, and two values of $D^{1/2} = 170$ MeV, $190$ MeV. These plots show again that the term $C$ can stiffen the EoS, pushing the occurrence of free quark phase deeper
into the star core, as discussed above. For the description of the NS's crust, we have joined the hadronic EoSs above described with the ones by Negele and Vautherin~\cite{negele:1973} in the medium-density regime ($0.001~$fm$^{-3}<\rho<0.08~$fm$^{-3}$), and the ones by Feynman-Metropolis-Teller~\cite{feynman:1949} and
Baym-Pethick-Sutherland~\cite{baym:1971} for the outer crust
($\rho<0.001$~fm$^{-3}$). 

Fig.~\ref{fig5} shows the corresponding HSs' masses as a function of their central densities (upper panel) and also the HS mass-radius
relations (lower panel) with the chosen parameters. The results of the nucleon star are also shown for comparison. We see that a higher mass is achieved when the term $C$ is included, for example, the value of $1.53 M_{\odot}$ ($1.65 M_{\odot}$) of the maximum mass obtained with $C = 0, D^{1/2} = 170$ MeV ($190$ MeV) is increased to $2.10 M_{\odot}$ ($2.17
M_{\odot}$) using $C = 0.7, D^{1/2} = 170$ MeV ($190$ MeV). Only in the case of $C = 0, D^{1/2} = 170$ MeV, a pure quark core can be
reached, while in other three cases, the most massive stars both have a mixed-phase core. We have also checked
that, an even larger $C$ value will bring heavier HSs, but always
with a mixed-phase core. This means that no pure core is possible
in the present HS model and quarks only appear in a limited region
of the NS's core. These results are consistent with a latest study
using the chiral effective field theory approach joined with the Polyakov$-$Nambu$-$Jona$-$Lasinio (PNJL)
model~\cite{weise14}.

\section{Conclusions}

Summarizing, we have presented updated calculations of the
transition from hadron to quark deconfined phase in NS matter, and
also the HS structure based on our previous work. This extension
concerns mainly the quark matter EoS, where we used a recent
derivation of the quark mass scaling, including the leading-order
perturbative interactions, in addition to the quark confinement.
The derivation scheme allows us to modify largely the high-density
behaviour of dense matter, resulting from a more repulsive quark
interaction.

We find that the quark thresholds are pushed to high densities,
together with  large density jumps in the first order phase
transition. Also, the EoSs are stiffened and the resulting HS
maximum mass are shifted to higher values. Massive HSs as high as
$2 M_{\odot}$ are possible, consistently with two recent
astrophysical observations of pulsars in binary systems. 

In the near future we plan to include the color superconductivity
since it is expected to play a role in dense quark matter at the
density range discussed in the present work. Also, the appearance
of hyperons, missing in our present version of BHF model, could be
studied in competition with free quarks, since a previous
study~\cite{dsm}, combining the same nucleon model with the
Dyson-Schwinger quark model shows that no hybrid star can exist if
hyperons are introduced. Finally, we would like to study how such
high-mass NSs are formed in a binary system.

\begin{acknowledgments}

We would like to thank Prof. D. Blaschke, Prof. X. J. Wen and Dr. C. J. Xia for valuable discussions. We also appreciate Prof. U. Lombardo for reading carefully our manuscript. The work was supported by the National Natural Science Foundation of China (Nos. 11135011, 11175219, 11435014, 11475110, U1431107), the Major State Basic Research Developing Program of China (No. 2007CB815004), and the Knowledge Innovation Projects (Nos. KJCX2-EW-N01, KJCX3-SYW-N2) of the Chinese Academy of Sciences.
\end{acknowledgments}



\begin{thebibliography}{99}

\bibitem{2times}
P. B. Demorest, T. Pennucci, S. M. Ransom, M. S. E. Roberts,
and J. W. T. Hessels, Nature (London) {\bf 467}, 1081 (2010); J.
Antoniadis, P. C. C. Freire, N. Wex, T. M. Tauris, R. S. Lynch
et al., Science {\bf 340}, 6131 (2013).

\bibitem{LOFT}
R. P. Mignani, S. Zane, D. Walton, T. Kennedy, B. Winter, P.Smith, R. Cole, D. Kataria, A. Smith [LOFT team Collaboration], arXiv:1201.0721.

\bibitem{rep}
 M. Prakash, I. Bombaci, M. Prakash, P. J. Ellis, J. M. Lattimer, and R. Knorren, Phys. Rep. {\bf 280}, 1 (1997).

\bibitem{Li08q}
G. X. Peng, A. Li, and U. Lombardo, Phys. Rev. C {\bf 77},
065807 (2008).

\bibitem{Li10k}
A. Li, X. R. Zhou, G. F. Burgio, and H.-J. Schulze, Phys. Rev. C {\bf 81}, 025806 (2010); A. Li, G. F. Burgio, U. Lombardo, and W. Zuo, Phys. Rev. C {\bf 74}, 055801 (2006); W. Zuo, A. Li, Z. H. Li, and U. Lombardo, Phys. Rev. C {\bf 70}, 055802 (2004).

\bibitem{Li11y}
G. F. Burgio, H.-J. Schulze, and A. Li, Phys. Rev. C {\bf 83}, 025804 (2011).

\bibitem{Li14y}
J. N. Hu, A. Li, H. Toki, and W. Zuo, Phys. Rev. C {\bf 89}, 025802 (2014).

\bibitem{book}
 M. Baldo,
 {\em Nuclear Methods and the Nuclear Equation of State},
 International Review of Nuclear Physics, Vol. 8
 (World Scientific, Singapore, 1999).

\bibitem{cubic}
G. X. Peng, H. C. Chiang, J. J. Yang, L. Li, and B. Liu,
Phys. Rev. C {\bf 61}, 015201 (1999).

\bibitem{horder}
C. J. Xia, G. X. Peng, S. W. Chen, Z. Y. Lu, and J. F. Xu, Phys. Rev. D {\bf 89}, 105027 (2014).

\bibitem{Aoki06}
Y. Aoki, G. Endrodi, Z. Fodor, S. D. Katz, and K. K. Szabo, Nature {\bf 443}, 675 (2006).

\bibitem{12}
N. Yasutake, T.  Noda, H. Sotani, T. Maruyama, and T. Tatsumi,
arXiv:1208.0427.

\bibitem{13}
M. Hempel, G. Pagliara, and J. Schaffner-Bielich,
Phys. Rev. D {\bf 80}, 125014 (2009).

\bibitem{Bla14}
D. Blaschke, D. E. Alvarez-Castillo, and S. Benic, arXiv:1402.0478; S. Benic, D. Blaschke, D. E. Alvarez-Castillo, T. Fischer, and S. Typel, arXiv:1411.2856.

\bibitem{Hat13}
K. Masuda, T. Hatsuda and T. Takatsuka, Prog. Theor. Exp. Phys. {\bf 7}, 073D01 (2013).

\bibitem{bhf}
 J.-P. Jeukenne, A. Lejeune, and C. Mahaux. Phys. Rep., {\bf 25C} (1976).

\bibitem{v18}
 R. B. Wiringa, V. G. J. Stoks, and R. Schiavilla,
 Phys. Rev. C {\bf 51}, 38 (1995).

\bibitem{Zuowei02prc}
W. Zuo, A. Lejeune, U. Lombardo, and J.-F. Mathiot,
 Nucl. Phys. A {\bf 706}, 418 (2002);
 Eur. Phys. J. A {\bf 14}, 469 (2002).

\bibitem{bbb}
 M. Baldo, I. Bombaci, and G. F. Burgio,
 Astron. Astrophys. {\bf 328}, 274 (1997).

\bibitem{both}
G. X. Peng, H. C. Chiang, and P. Z. Ning, Int. J. Mod. Phys.
A {\bf 18}, 3151 (2003).

\bibitem{liangmn}
A. Li, R. X. Xu, and J. F. Lu, Mon. Not. R. Astron. Soc. {\bf 402},
2715 (2010).

\bibitem{WenxjPRC72}
X. J. Wen, X. H. Zhong, G. X. Peng, P. N. Shen, and P. Z. Ning,
Phys. Rev. C {\bf 72}, 015204 (2005).

\bibitem{upqcd} S. Aoki, et al. preprint(hep-lat/13108555).

\bibitem{hybridbhf04}
C. Maieron, M. Baldo, G. F. Burgio, and H.-J. Schulze, Phys. Rev. D {\bf 70}, 043010 (2004).

\bibitem{1glue}
S. W. Chen and G. X. Peng, Commun. Theor. Phys. {\bf 57},
1037 (2012); Chin. Phys. C {\bf 36}, 947 (2012).

\bibitem{spinspin}
M. Oka and K. Yazaki, Phys. Lett. {\bf 90}, 41 (1980).

\bibitem{string}
 G.~R\"opke, D.~Blaschke and H.~Schulz,
  Phys. Rev. D {\bf 34}, 3499 (1986);  D.~Blaschke, T.~Tovmasian and B.~Kampfer,
  Sov. J. Nucl. Phys.  {\bf 52}, 675 (1990)
 [Yad. Fiz.  {\bf 52}, 1059 (1990)].

 \bibitem{shapiro}
 S. L. Shapiro and S. A. Teukolsky,
 {\em Black Holes, White Dwarfs, and Neutron Stars}
 (John Wiley and Sons, New York, 1983).

\bibitem{negele:1973} J. W. Negele, and D. Vautherin, Nucl. Phys. A
{\bf 207}, 298 (1973).

\bibitem{feynman:1949} R. Feynman, F. Metropolis, and E. Teller,
Phys. Rev. {\bf 75}, 1561 (1949).

\bibitem{baym:1971} G. Baym, C. Pethick, and D. Sutherland,
Astrophys. J. {\bf 170}, 299 (1971).

\bibitem{weise14}
T. Hell and W. Weise, Phys. Rev. C {\bf 90}, 045801 (2014).

\bibitem{dsm}
H. Chen, M. Baldo, G. F. Burgio, and H.-J. Schulze, Phys. Rev. D {\bf 84}, 105023 (2011).

\end{thebibliography}
\end{document}